\documentstyle[prl,aps,epsf,twocolumn]{revtex}
\newcommand{\la}{\langle}
\newcommand{\ra}{\rangle}

\newcommand{\beq}{\begin{eqnarray}}
\newcommand{\eeq}{\end{eqnarray}}

\newcommand{\btem}{\bibitem}

\begin{document}

\draft

\title{A Precursor of Chiral Symmetry Restoration in the Nuclear Medium}

\author{T. Hatsuda$^{(1)}$, T. Kunihiro$^{(2)}$, H. Shimizu$^{(3)}$}
\address{$^{(1)}$  Physics Department,Kyoto University, Kyoto 606-8502, Japan}
\address{$^{(2)}$  Faculty of Science and Technology, Ryukoku
 University, Seta, Otsu-city, 520-2194, Japan}
\address{$^{(3)}$  Physics Department, Yamagata University,
 Yamagata 990-8560, Japan}

\date{\today}
 
\maketitle

\begin{abstract}
Spectral enhancement  near the 2$m_{\pi}$ threshold  in the $I$=$J$=0 channel
in nuclei is shown to be a distinct  signal of the partial restoration of 
chiral symmetry. The relevance of this phenomenon with the possible detection
of  $2\pi^{0}$ and $2\gamma$  in hadron-nucleus and photo-nucleus reactions
 is discussed.
 \end{abstract}


One of the intriguing phenomena in the physics of strong interactions is the 
dynamical breaking of chiral  symmetry (DB$\chi$).
This explains the existence of the pion and also dictates
most of the low energy phenomena in hadron physics.
In the context of the Quantum Chromo Dynamics (QCD), the 
fundamental theory of strong interactions, DB$\chi$ 
is associated with the condensation of quark - anti-quark
pairs in the vacuum. This is analogous to the condensation of Cooper pairs 
in the theory of  superconductivity \cite{NJL}.
Furthermore, as the baryon density and/or the temperature are raised,
the QCD vacuum  is supposed to undergo a phase transition to the chirally
symmetric phase \cite{HK94}.
 In particular,  some theoretical analyses 
suggest that a partial restoration of chiral symmetry  occurs
even at a relatively low baryon density ($\rho$) close to  the 
nuclear matter density $\rho_0 = 0.17 {\rm fm}^{-3}$ \cite{HK94}.
Then, it is of fundamental importance
   to clarify what observables reflect most properly
this partial symmetry restoration.
The purpose of this Letter is to give a candidate of such  observables
 and to propose experiments to test the idea with nuclear targets.
 
The  general wisdom of many-body physics \cite{GK,dcf,nucl}  tells us that 
the  fluctuation of the order parameter becomes large as the 
system approaches the critical point of the phase transition.
In QCD, this corresponds to a softening of a collective excitation
having the same quantum number as that of the vacuum, 
namely the scalar-isoscalar ($I$=$J$=0) meson, $\sigma$ \cite{HK85}. 
The softening (the red-shift of the spectrum) in turn causes 
the  decrease of the decay width of $\sigma$
by the phase-space suppression of the reaction $\sigma \rightarrow 2 \pi$.
This leads to a conjecture that $\sigma$ may 
 appear as a sharp resonance at finite temperature ($T)$\cite{HK85,later},
although it can be  elusive due to the large width in the free 
space\cite{pipi,prim}.
A later analysis at $T\not=0$ showed that  
 the spectral function in the $\sigma$-channel 
 has a characteristic enhancement just above the
  two-pion threshold \cite{CH98}.

In this Letter, we demonstrate, using a toy model,
  that the spectral enhancement near the threshold is 
 a distinct signal of  the partial chiral restoration  
also at $\rho\not=0$.
   As possible experiments to observe this enhancement,
 we will propose to detect the neutral-dipion (2$\pi^0$)
 and diphoton (2$\gamma$)  in reactions with 
   heavy nuclear targets \cite{foot}.
 We will also mention the relevance of the softening to
  the recent data on the near-threshold
 $\pi^{+}\pi^{-}$ production in $\pi^{+}$-nucleus reactions
 by the CHAOS collaboration \cite{ppp}.

Before presenting our explicit model-calculation, let us
 describe the general
 features of the spectral enhancement near the two-pion threshold.
 Consider the propagator 
 of the $\sigma$-meson at rest in the medium :
$ D^{-1}_{\sigma} (\omega)= \omega^2 - m_{\sigma}^2$
$ - \Sigma_{\sigma}(\omega;\rho)$,
where $m_{\sigma}$ is the mass of $\sigma$ in the tree-level, and
$\Sigma_{\sigma}(\omega;\rho)$ is 
the loop corrections in the vacuum as well as in the medium.
The corresponding spectral function 
$\rho_{\sigma}(\omega) = - \pi^{-1} {\rm Im} D_{\sigma}(\omega)$
reads
\beq
\label{spect1}
\rho_{\sigma}(\omega) 
 = - {1 \over \pi} 
{ {\rm Im} \Sigma_{\sigma} \over (\omega^2 - m_{\sigma}^2 -
 {\rm Re} \Sigma_{\sigma} )^2
  + ({\rm Im} \Sigma_{\sigma} )^2 }.
\eeq
Near the two-pion threshold, the phase space factor gives
${\rm Im} \Sigma_{\sigma} \propto \theta(\omega - 2 m_{\pi}) \ 
	 \sqrt{1 - {4m_{\pi}^2 \over \omega^2}} $ in 
the one-loop order.
On the other hand, the partial restoration of chiral
symmetry indicates that $m_{\sigma}^*$ (the 
 ``effective mass'' of $\sigma$  defined through 
 ${\rm Re}D_{\sigma}^{-1}(\omega = m_{\sigma}^*)=0$)
approaches to $ m_{\pi}$.  Therefore,
 there exists a density $\rho_c$ at which 
 ${\rm Re} D_{\sigma}^{-1}(\omega = 2m_{\pi})$
 vanishes even before the complete $\sigma$-$\pi$
 degeneracy takes place; namely
 ${\rm Re} D_{\sigma}^{-1} (\omega = 2 m_{\pi}) =
 [\omega^2 - m_{ \sigma}^2 -
 {\rm Re} \Sigma_{\sigma} ]_{\omega = 2 m_{\pi}} = 0$.
At this point, the spectral function is solely dictated by the
 imaginary part of the self-energy;
\beq
\rho_{\sigma} (\omega \simeq  2 m_{\pi}) 
 =  - {1 \over \pi \ {\rm Im}\Sigma_{\sigma} }
 \propto {\theta(\omega - 2 m_{\pi}) 
 \over \sqrt{1-{4m_{\pi}^2 \over \omega^2}}}.
\eeq
This shows that, even if there is no sharp resonance
 in the scalar channel in the free space, there arises
  a mild (integrable) singularity just above the threshold in the medium.
We emphasize that this is a general phenomenon correlated with the 
 partial restoration of chiral symmetry.

 To make the argument more quantitative,
 let us evaluate $\rho_{\sigma}(\omega)$ in a {\em toy} model, namely
  the SU(2) linear $\sigma$-model:
\beq
\label{model-l}
{\cal L} & = &  {1 \over 4} {\rm tr} [\partial M \partial M^{\dagger}
 - \mu^2 M M^{\dagger} \\ \nonumber 
 & & \ \ \ \ \ \ \ \ \ \ \ \ \ \ \ \ \ 
  - {2 \lambda \over 4! } (M M^{\dagger})^2   - h (M+M^{\dagger}) ],
\eeq
where tr is for the flavor index and  
 $M = \sigma + i \vec{\tau}\cdot \vec{\pi}$.
  Although the model 
 is not a precise low energy representation of QCD \cite{GL}, it
 is known to describe the
 pion dynamics qualitatively well up to 1GeV as shown by
 Chan and Haymaker \cite{BW}.

 The coupling constants $\mu^2, \lambda$ and $h$ have
 been determined in the
 vacuum to reproduce $f_{\pi}=93$ MeV, $m_{\pi}=140$ MeV as well as
 the s-wave $\pi$-$\pi$ scattering phase shift in the one-loop order.
 Resultant parameters in the $\overline{MS}$ renormalization
 scheme are given  in \cite{CH98} and are recapitulated for two 
 characteristic cases in TABLE I in which
  $m_{\sigma}^{peak}$ is defined as 
  a peak position of $\rho_{\sigma}(\omega)$.

 The interaction Lagrangian of $M$ with the nucleon field  $N$
 with SU(2) chiral symmetry  is modeled as
\beq
\label{int-nm}
{\cal L}_{I}(N, M) =
  - g \chi \bar{N} U_5 N  - m_0  \bar{N} U_5 N ,
\eeq
where we have used a polar representation 
 $ \sigma + i  \vec{\tau}\cdot \vec{\pi} \gamma_5 \equiv \chi U_5 $ 
 for convenience \cite{esb}. 
 The first term in (\ref{int-nm}) with the coupling constant $g$
 is a standard chiral invariant coupling in the linear $\sigma$
 model.  The second term with a new parameter $m_0$, which is 
 usually not taken into account in the literature,
 is also chiral invariant and non-singular.

With the dynamical breaking of chiral symmetry
($\langle \sigma \rangle_{\rm vac} \equiv \sigma_0 \neq 0$), 
 eq.(\ref{int-nm}) expanded in
 terms of $\sigma/\sigma_0$ and $\vec{\pi}/\sigma_0$ reads,
${\cal L}_{I}(N, M) =  - m_N \bar{N} N 
 - \bar{N} (g_{\rm s} \tilde{\sigma} 
 + i g_{\rm p}  \vec{\tau}\cdot \vec{\pi} \gamma_5) N 
 + {1 \over 2} (m_0 / \sigma_0^2 ) \bar{N} \vec{\pi}^2 N 
 + 0(\tilde{\sigma}^{l \ge 1} \times \pi^{n \ge 1} )$,
 where $\tilde{\sigma} = \sigma - \sigma_0 $,
 $m_N \equiv m_0 + g \sigma_0$, $g_{\rm s} \equiv g$ and   
 $g_{\rm p} \equiv g_{\rm s} + m_0 /  \sigma_0$.
   Because of $m_0$,
 the standard constraint $g_s = g_p$ can be relaxed
 without conflicting with chiral symmetry.
 Also, the term proportional to $m_0 \pi^2$ appears to
 preserve chiral symmetry. 
   $g_{\rm p}$ is constrained by 
 the Goldberger-Treiman relation; 
 $g_{\rm p} = m_N /  \sigma_0 \simeq m_N/f_{\pi}
 = g_{\pi N} = 13.5$.  On the other hand,
 $g_{\rm s}$ is independent of $g_{\rm p}$
 and can be treated as a free parameter.
 With this freedom, one can circumvent the well-known
 problem  that 
  $g_s=g_p$ combined with eq.(\ref{model-l})
 does not reproduce the known nuclear
 matter properties in the mean-field level \cite{sw}.
 We remark that the dilated chiral model \cite{BV}
 can also avoid  $g_s = g_p$ away from
 the ``dilaton'' limit,  and has been applied
 to study the scalar meson in nuclear matter \cite{KL}.

 In the following, we treat the effect of the meson-loop as well as
 the baryon density  as a perturbation 
 to the vacuum quantities. Therefore, our loop-expansion  
  is valid only at relatively low
 densities. 
 The full self-consistent
 treatment of the problem requires  systematic resummation of loops
 similar to what was developed at finite $T$ \cite{CH98}. 
 Let us first consider the chiral condensate in nuclear matter
 $\langle \sigma \rangle$ and parametrize it as
\beq
\label{cond}
\langle \sigma \rangle \equiv  \sigma_0 \ \Phi(\rho).
\eeq
In the linear density approximation,
 $\Phi(\rho) = 1 - C \rho / \rho_0 $ with
 $C = (g_{\rm s} /\sigma_0 m_{\sigma}^2) \rho_0$.
 Instead of using $g_{\rm s}$, we
   use $\Phi$  as a basic parameter in the 
 following analysis.
  The plausible value of $\Phi(\rho = \rho_0)$ is
 0.7 $\sim$ 0.9 \cite{HK94}.

The one-loop corrections to the self-energy for $\sigma$ can be read off
 from the  diagrams in Fig.1:
$\Sigma_{\sigma}(\omega;\rho) = \Sigma_{\rm vac}^{A} $
$ + \Sigma_{\rm vac}^B + \Sigma_{MF}(\rho) + \Sigma_{ph}(\rho)$.
 $\Sigma_{\rm vac}^A$ ($\Sigma_{\rm vac}^B$) corresponds to
 Fig.1A and Fig.1B respectively. Only the latter has  
 $\omega$-dependence and  the imaginary part.
  The explicit formula for
  $\Sigma_{\rm vac}^{A+B}$  renormalized in the
 $\overline{MS}$ scheme is given in Appendix A of \cite{CH98}.

 $\Sigma_{MF}(\rho)$ corresponds to the mean field correction
 in the nuclear matter (Fig.1C).
 $\Sigma_{ph}(\rho)$ is a correction from the nuclear particle-hole 
 excitation. We take only the density dependent part
 in these diagrams and
 neglect the problematic vacuum-loops of the nucleon 
 \cite{sw}.
 
The leading term in the mean-field part is easily estimated as
\beq
\Sigma_{MF}(\rho) =  \lambda \sigma_0 \ (\la \sigma \ra - \sigma_0)
 = - \lambda \sigma_0^2 \ (1-\Phi(\rho)) ,
\eeq
 Leading term  in the particle-hole part (Fig.1D)
 in terms of $k_F = (3 \pi^2 \rho /2)^{1/3}$ reads
 $\Sigma_{ph}(\rho)   \simeq  {2 g_s^2 \over 5 \pi^2} 
{k_F^5 \over M^3}$, 
 which starts from $O(\rho^{5/3})$
  and is not more than a few \% of $\Sigma_{MF}(\rho)$
  at  $\rho = \rho_0$.
 This is in contrast to the case of the pion, where
 both $\Sigma_{MF}(\rho) $ and $\Sigma_{ph}(\rho) $ are
 proportional to $\rho$ and cancel with each other due to chiral symmetry.  
 Because of this cancellation, we can neglect the 
 two-loop contribution related to the medium modification
 of  the low-momentum pions  near
 the 2$m_{\pi}$ threshold.

 Up to this order,
 ${\rm Im} \Sigma_{\sigma}$  solely comes from 
${\rm Im} \Sigma_{\rm vac}^B$, since 
 there is no Landau damping and scalar-vector
 mixing for the $\sigma$-meson at rest in nuclear matter:
\beq
\label{vac4}
{\rm Im} \Sigma_{\sigma}(\omega;\rho) =
{\rm Im} \Sigma_{\rm vac}^{B}
 = - {\lambda^2 \over 96 \pi}  \sigma_0^2
 \sqrt{1 - {4m_{0\pi}^2 \over \omega^2}} \  ,
\eeq
for $2 m_{\pi} \le \omega \le 2m_{\sigma}$.

Now, let us look at the spectral function defined 
 in (\ref{spect1}).
As we have already discussed,
the threshold peak is prominent when
  ${\rm Re} D_{\sigma}^{-1} = \omega^2 - m_{ \sigma}^2 -
 {\rm Re} \Sigma_{\sigma}  =0 $. 
 In the parametrization given in (\ref{cond}),
 this condition is  rewritten as
 $\Phi(\rho_c) = 0.74$ \ \ \ (case  (I)), and 
     $\Phi(\rho_c) = 0.76$  \ \ \ (case  (II)).
 The numbers in the right hand side are insensitive to the
 parameters  in eq.(\ref{model-l})
 as far as the physical quantities in the vacuum such as
 $f_{\pi}$ and $m_{\pi}$ are fixed.
 In the linear density formula $\Phi(\rho)=1-C\rho/\rho_0$ with 
 a reasonable value  $C \simeq 0.2$, we obtain
  $\rho_c \simeq 1.25 \rho_0$ which is not far from the normal nuclear
 matter density.
 This implies that 
 we could see the threshold  enhancement in 
 experiments with heavy nuclear targets.

The spectral functions 
 together with ${\rm Re} D_{\sigma}^{-1}(\omega)$  
 for two cases (I) and (II) 
 are shown in Fig.2 and in Fig.3, respectively.
   In both figures, the characteristic enhancement 
 just above the 2$m_{\pi}$  threshold is seen for
  $\rho \simeq \rho_c$  \cite{sum}.
 
We notice that the enhancement is caused by (a) partial restoration
 of chiral symmetry where $m_{\sigma}^*$ approaches toward
 $m_{\pi}$,  
 and (b) the cusp structure of 
${\rm Re}D_{\sigma}^{-1}(\omega)$ at $\omega = 2 m_{\pi}$; see the 
 lower panels of Fig.2 and Fig.3.
 Although the cusp is not prominent at zero density,
 it eventually hits the real axis at $\rho = \rho_c$
 because ${\rm Re}D_{\sigma}^{-1}(\omega )$ increases associated
 with $m_{\sigma}^* \rightarrow 2 m_{\pi}$.
  This is a general phenomenon for systems where the 
 internal symmetry is partially restored in the medium \cite{su}.
  Another important observation is that, 
  even at  densities 
 well below the point where $m_{\sigma}^*$ and $m_{\pi}$ are
 degenerate,
 one can expect  the spectral enhancement near 
the $2m_{\pi}$ threshold \cite{quest}.

To confirm the threshold enhancement associated with the
 partial chiral restoration, measuring 2$\pi^0$ and 
$2\gamma$ in experiments with hadron/photon beams off
 the  heavy nuclear targets should be most appropriate. 
 Measuring $\sigma \rightarrow 2 \pi^0 \rightarrow
  4\gamma$ is experimentally feasible 
 \cite{4gamma},  and
 one can avoid the possible $I=J=1$ background from the
 $\rho$ meson inherent in the $\pi^+\pi^-$ measurement.
 Measuring the direct electromagnetic decay $\sigma \rightarrow 2 \gamma$
 is also important because of the small final state
 interactions. However, 
 the branching ratio is small in this case
 (Br $= \Gamma_{\sigma \rightarrow 2 \gamma} 
  / \Gamma_{\sigma \rightarrow 2 \pi} = O(10^{-5})$ \cite{pipi}), and 
 one needs to fight with large 
 background of photons mainly coming from $\pi^0$'s.
 Nevertheless,  if the enhancement
  is prominent and changes rapidly as the mass number of the target nucleus,
  there is a chance to find the signal.  
 There is also a possibility that one can detect dilepton 
 through the scalar-vector mixing in matter: $\sigma \to \gamma^* \to
 e^+ e^-$ \cite{lepton}. In this case,
  the dileptons are produced  only when
 $\sigma$ has a finite three momentum.

 To enhance the production cross section for the 
 critical fluctuation in the $I$=$J$=0 channel,
   (d, $^3$He)  reactions is useful.
 The incident kinetic energies of the
  deuteron in the laboratory
 system $E$ can be estimated to be
  $1.1 {\rm GeV} < E < 10$ GeV, 
  to cover the spectral function 
 in the range  $2m_{\pi} < \omega < 750$ MeV \cite{HHG}.

Recently  CHAOS collaboration  \cite{ppp} measured the 
$\pi^{+}\pi^{\pm}$
invariant mass distribution $M^A_{\pi^{+}\pi^{\pm}}$ in the
 reaction $A(\pi^+, \pi^{+}\pi^{\pm})X$ with the 
 mass number $A$ ranging
 from 2 to 208: They observed that
the   yield for  $M^A_{\pi^{+}\pi^{-}}$ 
 near the 2$m_{\pi}$ threshold is close to zero 
for $A=2$, but increases dramatically with increasing $A$. They
identified that the $\pi^{+}\pi^{-}$ pairs in this range of
 $M^A_{\pi^{+}\pi^{-}}$ is in the $I=J=0$ state.
  Attempts so far in hadronic models without considering the
 partial chiral restoration 
 failed to reproduce this enhancement \cite{wambach,oset}.
 On the other hand, 
 the invariant mass distribution presented in \cite{ppp} 
 near 2$m_{\pi}$ threshold for large $A$ has a close
 resemblance to our model calculation in Fig.2, which suggests
 that this experiment may already provide
  a hint about how the chiral symmetry
 is (partially) restored at finite density.

In summary,  we have shown that the spectral function in the
 $I=J=0$ channel has 
 a large enhancement near the $2m_{\pi}$ threshold even at
  nuclear matter density due to the partial chiral restoration.
  Detection of the 
 dipion and diphoton spectral distribution
 in the reactions of hadron/photon with heavy nucleus is suitable
  to confirm the idea of partial chiral restoration in nuclei.

\vspace{0.5cm}

This work was initiated when T. H. and T.K. attended
  GSI  workshop on Hadrons in Matter (July 13, 1998).
 They are grateful to B. Friman, W. N\"{o}renberg,
 and the theory department at GSI for their invitation
 and warm hospitality. 
 T.K. thanks H. Pirner, C. Wetterich and the members of 
 Institute for Theoretical Physics, 
 Heidelberg University for their hospitality.
 He also thanks DAAD to make his  stay in Germany possible.

{\bf Note added:}
  After the completion of this paper, we got aware of a very recent 
 study on the CHAOS data; 
 R. Rapp et al., {\tt nucl-th/9810007}.
 They  do not  address the question of the
 partial chiral restoration, and the underlying medium effect responsible for
 the near-threshold enhancement  is 
  different from ours.

\begin{table}[h]
\begin{tabular}{|c|c|ccc|} 
                            &
$m_{\sigma}^{peak}$ (MeV) \ \ &
$\sqrt{-\mu^2}$ (MeV)         &
$\lambda/4 \pi$              & 
$h^{1/3}$  (MeV)                  \ \\ \hline
(I)  &  550 & 284 & 5.81  & 123   \\
(II) &  750 & 375 & 9.71  & 124   \\
\end{tabular}
\vskip4mm
\caption{Parameters for $m_{\sigma}^{peak}=$ 550, 750 MeV} 
\label{tab1}
\end{table} 
%
\begin{figure}
\epsfxsize=7cm
\centerline{\epsfbox{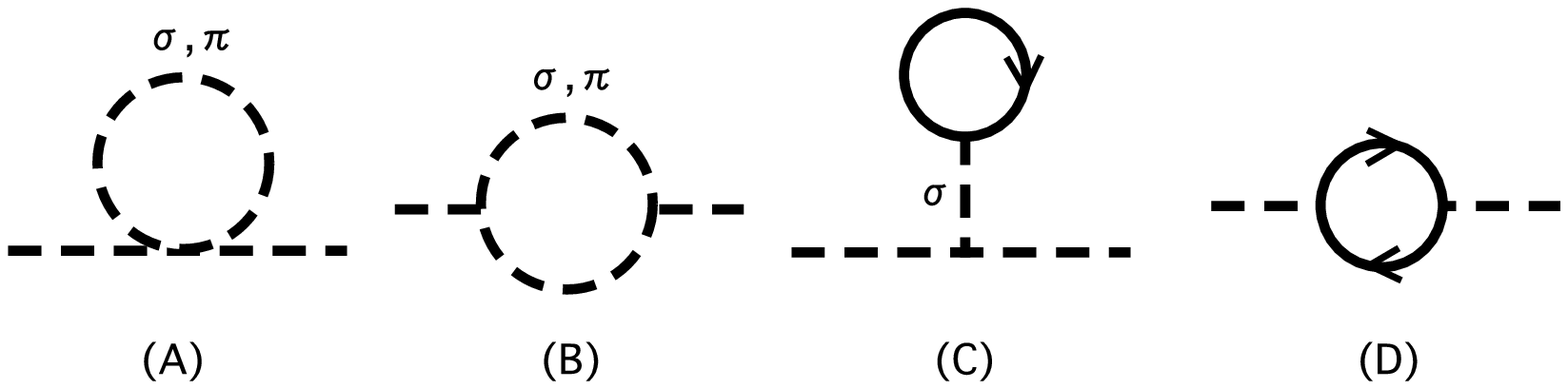}}
\vskip4mm
\caption{One-loop self energy.
 The dashed line denotes either $\sigma$ or $\pi$.
 The solid line denotes the nucleon.  }
\label{fig1}

\vskip4mm
\epsfxsize=6.5cm
\centerline{\epsfbox{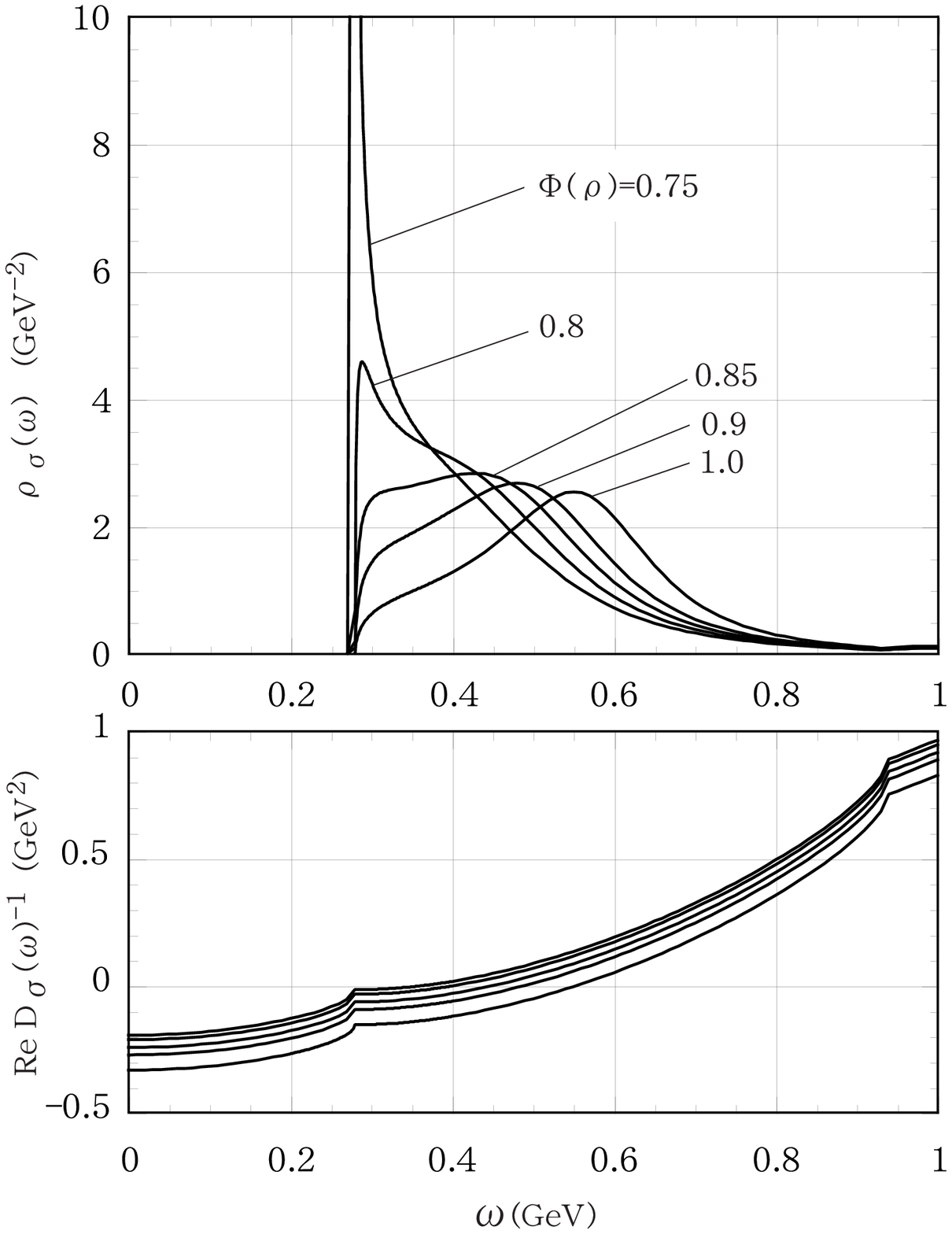}}
\vskip4mm
\caption{Spectral function for $\sigma$ and  the 
 real part of the inverse propagator for several values of
 $\Phi = \la \sigma \ra / \sigma_0$ with $m_{\sigma}^{peak}
 = 550$ MeV (case (I) in TABLE I). In the lower panel,
 $\Phi$ decreases from bottom to top.}
\label{fig2}

\vskip4mm
\epsfxsize=6.5cm
\centerline{\epsfbox{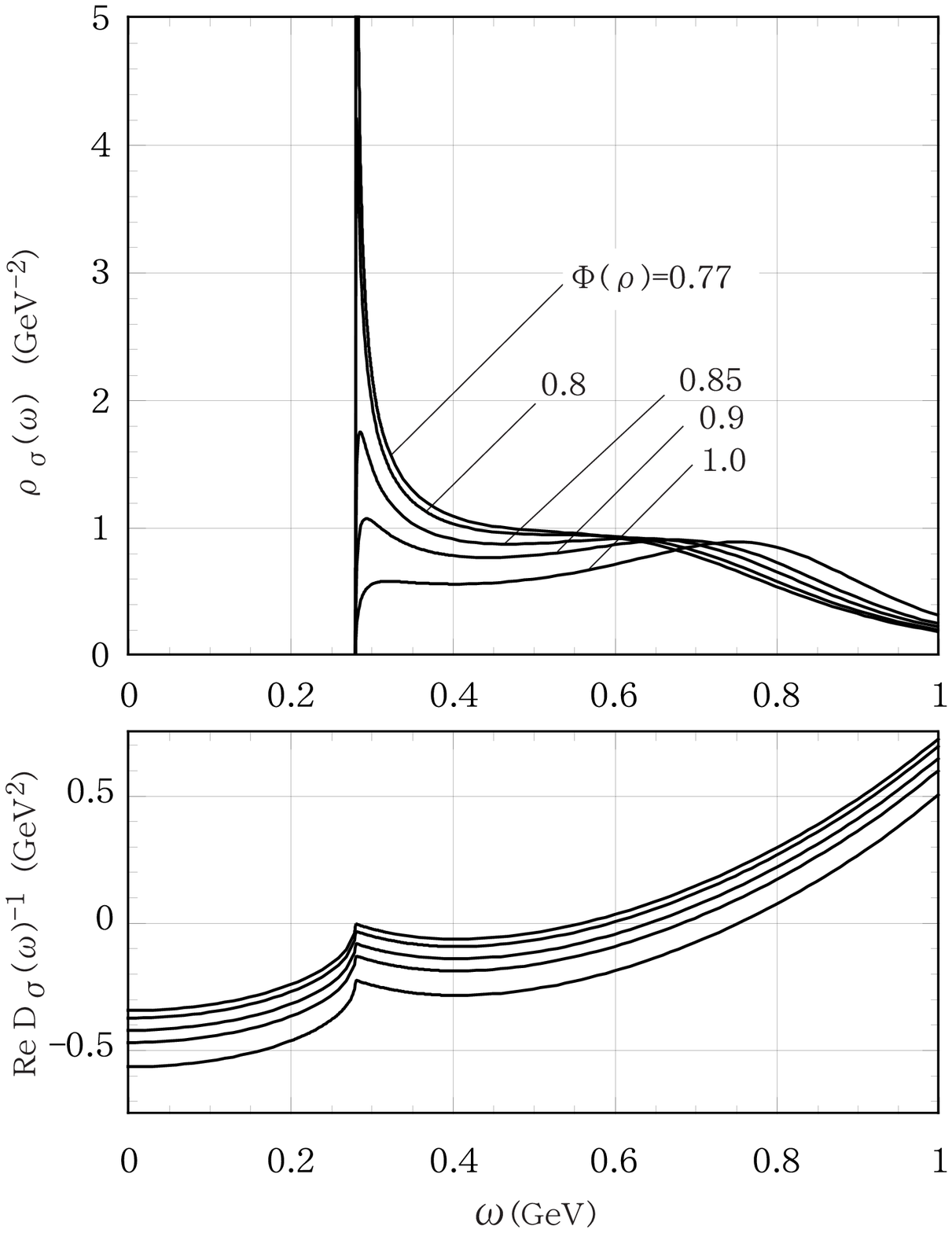}}
\vskip4mm
\caption{Same with Fig.2 for $m_{\sigma}^{peak}= 750$ MeV 
(case (II) in TABLE I.}
\label{fig3}

\end{figure}

\end{document}